\documentclass[aps,prd,nofootinbib,twocolumn,superscriptaddress,preprintnumbers,letterpaper,natbib,nofootinbib]{revtex4-2}

\pdfoutput = 1
\usepackage{amssymb}
\usepackage{amsmath}
\usepackage{epsfig}
\usepackage{hyperref}
\usepackage{color}
\usepackage{cleveref}
\usepackage{multirow}
\usepackage{capt-of}
\usepackage{siunitx}
\usepackage{graphicx}
\usepackage[caption=false]{subfig}
\usepackage{slashed}
\usepackage{tabularx}
\usepackage{enumitem}
\usepackage{multirow}
\usepackage{comment}
\usepackage{url}
\usepackage{xspace}
\usepackage{hhline}

\newcommand{\bea}{\begin{eqnarray}}
\newcommand{\eea}{\end{eqnarray}}

\makeatletter
\def\simgt{\mathrel{\lower2.5pt\vbox{\lineskip=0pt\baselineskip=0pt
           \hbox{$>$}\hbox{$\sim$}}}}
\def\simlt{\mathrel{\lower2.5pt\vbox{\lineskip=0pt\baselineskip=0pt
           \hbox{$<$}\hbox{$\sim$}}}}
\makeatother

\newcommand\psiB{\psi_{\mathcal{B}}}
\newcommand\psiBbar{\bar{\psi}_{\mathcal{B}}}

\newcommand{\mysec}[1]{\noindent {\textit{#1.}}---}

\newcommand{\Y}{\mathcal{Y}}
\newcommand{\M}{3\emph{M}}
\newcommand{\Br}{\text{Br}(B^0\rightarrow \psiBbar  \,\mathcal{B}_{\rm SM})}
\newcommand{\Bri}{\text{Br}\left(B_i^0\rightarrow \psiBbar \, \mathcal{B}_{\rm SM}\right)}

\begin{document}

\preprint{CERN-TH-2024-144}
\preprint{UT-WI-28-2024}

\title{
The Standard Model CP Violation is Enough
}

\author{Gilly Elor}
\email{gilly.elor@austin.utexas.edu}
\affiliation{Weinberg Institute, Department of Physics, University of Texas at Austin, Austin, TX 78712, United States}

\author{Rachel Houtz}
\email{rachel.houtz@ufl.edu}
\affiliation{Department of Physics, University of Florida, Gainesville, FL 32611, USA}
\affiliation{Theoretical Physics Department, CERN, 1211 Geneva, Switzerland}

\author{Seyda Ipek}
\email{Seyda.Ipek@carleton.ca}
\affiliation{Carleton University  1125 Colonel By Drive, Ottawa, Ontario K1S 5B6, Canada}

\author{Martha Ulloa}
\email{m.ulloacalzonzin@ufl.edu}
\affiliation{Department of Physics, University of Florida, Gainesville, FL 32611, USA}

\begin{abstract}
Is the Standard Model Charge-Parity (CP) violation ever enough to generate the observed baryon asymmetry? Yes! We introduce a mechanism of baryogenesis (and dark matter production) that can generate the entire observed baryon asymmetry of the Universe using \emph{only} the CP violation within Standard Model systems--- a fête which no other mechanism currently proposed can achieve. Baryogenesis proceeds through a Mesogenesis scenario but with well motivated additional dark sector dynamics: a \emph{morphon} field generates present day mass contributions for the particle mediating the decay responsible for baryogenesis. The effect is an enhancement of baryon production whilst evading present day collider constraints. The CP violation comes entirely from Standard Model contributions to neutral meson systems. Meanwhile, the dark dynamics generate gravitational waves that may be searched for with current and upcoming Pulsar Timing Arrays, as we demonstrate with an example.  This mechanism, \emph{Mesogenesis with a Morphing Mediator}, motivates probing a new parameter space as well as improving the sensitivity of existing Mesogenesis searches at hadron and electron colliders. 
\end{abstract}

\maketitle

\mysec{Introduction}
The quest to discover the theory that creates the baryon asymmetry of the Universe (BAU) has been ongoing since Sakharov first laid out the three conditions needed to generate a primordial asymmetry of matter over antimatter \cite{Sakharov:1967dj}: the breaking of baryon number, violation of charge (C) and charge-parity (CP) symmetries, and the existence of out-of-equilibrium interactions. While aspects of the Sakharov criteria  exist within the Standard Model (SM), it is widely appreciated that beyond the SM physics is required to successfully generate the observed BAU. Deviating from the traditional approach of all-encompassing new physics, in this letter we introduce the \emph{only} mechanism, to date, which succeeds in generating the BAU with \emph{just} the SM CP violation (CPV).

CPV arises in the SM as an irreducible phase in the quark mixing matrix \footnote{A possible CPV phase in the neutrino mixing matrix has yet to be experimentally confirmed.}. While this CPV phase is not small, the parametrization-independent measure of SM CPV, namely the Jarlskog parameter~\cite{Jarlskog:1985ht} is $J\sim 10^{-5}$. In electroweak baryogenesis models, where this CPV \emph{could} be relevant, CPV effects are further suppressed by small Yukawa couplings resulting in an asymmetry that is 10 orders of magnitude smaller than the measured value~\cite{,Gavela:1993ts, Gavela:1994ds, Gavela:1994dt,Huet:1994cr,Huet:1994jb}. One approach has been to utilize the CKM phase by making the Yukawa couplings larger in the early universe ~\cite{Berkooz:2004kx, Perez:2005yx,Bruggisser:2017lhc} \footnote{Note that \cite{Bruggisser:2017lhc} requires extra CP violation.}. However, such large Yukawa couplings destabilize the Higgs potential~\cite{Braconi:2018gxo}. 

The SM CPV also arises in neutral meson oscillations. In this letter, we exploit the CPV in the neutral $B$-meson system and show that it can be enough to generate the entire BAU. One way this scenario can be realized is through  \emph{Mesogenesis}~\cite{Elahi:2021jia, Elor:2020tkc, Elor:2018twp}. In Mesogenesis, SM mesons, that are produced out-of-equilibrium (when the Universe was at a temperature of  $5-100$~MeV), decay into a dark sector states with SM baryon number. Various Mesogenesis mechanisms leverage the CPV in charged and neutral meson systems, and the BAU is directly linked to experimental observables \cite{Berger:2023ccd,Elor:2022jxy, Alonso-Alvarez:2021qfd, Borsato:2021aum,Alonso-Alvarez:2021oaj}, e.g the  semi-leptonic asymmetry in neutral meson oscillations --- which are computed to be sizeable even without new physics concatenations to the SM \cite{Lenz:2019lvd}. However, the SM CPV alone is still not enough \cite{Alonso-Alvarez:2021qfd}.

Mesogenesis mechanisms, by construction, require a dark sector which shares an equal and opposite baryon asymmetry to that of the SM sector. There must also exist a SM-dark sector mediator $\Y$ which allows SM mesons to decay into the dark sector. In doing so, Mesogenesis also provides an explanation for the nature and origin of dark matter. In this letter we consider various dark sector dynamics which allow the SM-dark sector mediator $\Y$ to acquire a temperature dependent contribution to its mass. Using this simple, yet theoretically well motivated, \emph{morphing} of the mediator mass, we show that there does exists a mechanism of baryogenesis where the SM CPV is enough. Furthermore, the dark sector dynamics open up  new classes of signals which augment mnk the existing Mesogenesis experimental program --- with unique signals of \emph{Mesogenesis with a Morphing Mediator} (\M).

\medskip
\mysec{Mesogenesis with a Morphing Mediator}
Mesogenesis requires interactions between SM quarks and a new dark fermion $\psiB$ carrying anti-baryon number, $B=-1$, giving rise to a ($B$-conserving) effective operator: 
\bea
\mathcal{O}_{d_k, u_i d_j} = \mathcal{C}_{d_k, u_i d_j} \epsilon_{\alpha \beta \gamma} (\psiBbar d_k^\alpha ) (\bar{d}_j^{c \beta} u_i^\gamma)  \,,
\label{eq:op}
\eea
$i,j$ and $\alpha, \, \beta, \, \gamma$ are flavor and color indices respectively. Eq.~\eqref{eq:op} can arise in a UV model with  
a heavy color triplet scalar $\Y$. Assigning $\Y$ an electric charge $-1/3$ and baryon number $-2/3$ \footnote{Other choices are possible but are relatively inconsequential to the final results.}, symmetry admits the following Lagrangian:
\bea
\hspace{-0.18in} \mathcal{L}_{\Y} =  -\sum_{i,j} y_{u_i d_j} \Y^\star \bar{u}_{iR} d_{jR}^c - \sum_k y_{\psi d_k} \psiBbar \Y d_{kR}^c  + \text{h.c.}
\label{eq:UVmodel}
\eea
In a simple supersymmetric realization \cite{Alonso-Alvarez:2019fym}  $\Y$ is identified with a squark. LHC searches for squarks constrain the Wilson coefficient of the effective operators: $\mathcal{C}_{d_k, u_i d_j}  \equiv y_{\psi d_k} y_{u_i d_j} /M_\Y^2$, where the couplings $y_{\psi d_k} $,   $y_{u_i  d_j} \lesssim 4 \pi$ by perturbativity. Hence the $\Y$ mass, as with squarks, is constrained to be heavier than a few TeV \cite{Alonso-Alvarez:2021qfd}. Proton decay through $\psiB u dd$ is kinematically forbidden by requiring all dark sector baryons to have a mass $\gtrsim 1 \, \text{GeV}$.   

\begin{table}[t!]
\renewcommand{\arraystretch}{1.25}
  \setlength{\arrayrulewidth}{.25mm}
\centering
\small
\setlength{\tabcolsep}{0.18 em}
\begin{tabular}{ |c | c | c | c |  }
\hline
    Operator &  $\,\,(M^f_\Y)_{\rm min} \left[ \text{TeV} \right]\,\,$  & Decay  &  $\,\,\,\,\,\, \Gamma_0 \left[ \text{GeV}^{5} \right]\,\,\,\,\,\,$         \\
\hline
\hline
$\mathcal{O}_{b,ud}$ &  $\sim 1.7\sqrt{y_{\psi b}  \, y_{u d}}$  & $B_d \rightarrow \psiBbar \, n$ & 
 $3.5_{\pm 0.4} \! \cdot \! 10^{-5}$  \\
\, & &  $\,\,\,\, B_s \rightarrow \psiBbar \, \Lambda \,\,\,\,$ & n.a.  \\
\hline
 $\mathcal{O}_{b,us}$  &  $\sim 1.7\sqrt{y_{\psi b}  \, y_{us}}$ &  $\,\,B_d \rightarrow \psiBbar \, \Lambda \,\,$ & $1.4_{\pm 0.1} \! \cdot \! 10^{-4}$  \\  
\, & &  $B_s \rightarrow \psiBbar \,\Xi^0$ &  $3.2_{\pm 0.1} \!\cdot \! 10^{-5}$ \\
\hline
$\mathcal{O}_{b,cd}$ & $\sim  0.9 \sqrt{y_{\psi b}  \, y_{c d}}$  & $B_d \rightarrow \psiBbar \, \Sigma_c^0$ & $0.7_{\pm 0.4} \!\cdot \! 10^{-6}$ \\
\,& &  $B_s \rightarrow \psiBbar \, \Xi_c^0$& $6.6_{\pm 3.3} \!\cdot \! 10^{-7}$ \\
\hline
 $\mathcal{O}_{b,cs}$ & $\sim 0.9 \sqrt{y_{\psi b}  \, y_{c s}}$  & $B_d \rightarrow \psiBbar \, \Xi_c^0$ &$4.7_{\pm 2.0} \! \cdot \! 10^{-6}$  \\
\, & &  $B_s \rightarrow \psiBbar \, \Omega_c$ &  $5.0_{\pm 3.0} \!\cdot \! 10^{-6}$ \\ 
\hline

\end{tabular}
\caption{For operators $\mathcal{O}_{b,u_i d_j} \equiv i \epsilon_{\alpha \beta \gamma} b^\alpha (\bar{d}_j^{c \beta} u_i^\gamma)$, 
we quote the lower bound on $\Y$ mass, as constrained by present day collider searches ~\cite{Alonso-Alvarez:2021qfd}. Each operator mediates $B_{s,d}$ decays to various final baryons and missing energy ($\psiB$). We quote the maximal value (evaluated at minimum possible $\psiB$ mass i.e. 1 GeV) of this decay rate \cite{Elor:2022jxy}, with coefficients factored out: $\Gamma_0 \equiv \Gamma_B |_{m_{\psiB}=1\text{GeV}} / \mathcal{C}_{b,u_i d_j}^2$ .
}
\label{tab:decays}
\end{table}

In Neutral $B$-Mesogenesis \cite{Elor:2018twp}, equal numbers of neutral $B_{s,d}$ mesons and anti-mesons are produced by the late-time decay of a heavy scalar $\Phi$ \footnote{This could be a scalar field in a multi-field inflation scenario, a moduli field, or even a saxion; the details have little to no consequences on baryogenesis, dark matter, or the results of this work.}  at a temperature $T_d\lesssim 100 \text{MeV}$, and undergo CP violating meson-antimeson oscillations. Next, portal operators in Eq. ~\eqref{eq:op} involving one $b$-quark mediate the out-of-thermal equilibrium  meson decay: $B^0_{d,s} \rightarrow \bar{\psi}_{\mathcal{B}} \, \mathcal{B}_{\rm SM}$, into the dark anti-baryon and a SM baryon. The result is an equal and opposite baryon asymmetry between the dark and SM sectors. $\psiB$ ultimately decays into the stable dark matter of the Universe. The nature of the final state SM baryon depends on the flavorful variations \footnote{There is no \emph{a priori} reason to assume a specific flavor structure \cite{Alonso-Alvarez:2019fym, 
Alonso-Alvarez:2021qfd}.} of Eq.~\ref{eq:op}. Four different operators, $\psiB bus$, $\psiB bud$, $\psiB bcs$ and $\psiB bcd$, can generate the BAU. 

Mesogenesis occurs when the temperature of the Universe, $T_d$, was $O(10~{\rm MeV})$; before the era of Big Bang Nucleosythasis (BBN) but after the quark-hadron phase transition. The measured BAU, defined as the baryon-to-entropy ratio $Y_{\mathcal{B}}^{\rm meas} \equiv (n_{\mathcal{B}} - n_{\bar{\mathcal{B}}})/s = (8.718\pm 0.004) \times 10^{-11}$ \cite{Ade:2015xua,Aghanim:2018eyx,Cyburt:2015mya,pdg}, is generated as \cite{Elor:2018twp, Alonso-Alvarez:2019fym, Alonso-Alvarez:2021qfd}:
\bea
\!\!\!\! Y_{\mathcal{B}} \simeq 5 \times 10^{-5} \, \sum_{i=d,s} \bigl[ \Bri  A_{sl}^i \bigr]  \alpha_i (T_d) \,,
\label{eq:YBmeso}
\eea
where the inclusive branching fraction is over all possible final states. $A_{ sl}^{i=s,d}$ is the semi-leptonic asymmetry, an \emph{observable} CP violating parameter in  $B^0$ meson oscillations, $A_{sl}^i \equiv \frac{\Gamma \left( \bar{B}_i^0 \rightarrow B_i^0 \rightarrow f\right) - \Gamma\left(\bar{B}_i^0  \rightarrow B_i^0 \rightarrow \bar{f}\right) }{\Gamma\left( \bar{B}_i^0  \rightarrow B_i^0\rightarrow f\right) + \Gamma \left( \bar{B}_i^0  \rightarrow B_i^0 \rightarrow \bar{f}\right) }$, for various final states $f$. The $T_d$ dependent functions $\alpha_i \in \left[0,1 \right]$ capture additional numerics and higher temperature decoherence effects that spoil $B^0$-$\bar{B}^0$ oscillations (and thus the generated BAU) \cite{Elor:2018twp}. $\alpha_{s,d}$ peak at $T_d=20$ MeV and 10 MeV, respectively \cite{Alonso-Alvarez:2019fym, Alonso-Alvarez:2021qfd}.
Generically, $B^0_s$ mesons dominate production at higher $T_d > 35$ MeV while $B_d$ mesons are more relevant at lower $T_d$. Dependence on the parent particle, $\Phi$, parameters is weak in the regions we consider.

\begin{table}[t!]
\renewcommand{\arraystretch}{1.25}
  \setlength{\arrayrulewidth}{.25mm}
\centering
\small
\setlength{\tabcolsep}{0.18 em}
\begin{tabular}{ |c | c | c | c |  }
\hline
    Operator &  $\,\,(M^f_\Y)_{\rm min} \left[ \text{TeV} \right]\,\,$  & Decay  &  $\,\,\,\,\,\, \Gamma_0 \left[ \text{GeV}^{5} \right]\,\,\,\,\,\,$         \\
\hline
\hline
$\mathcal{O}_{d,ub}$ & $\sim 3.8 \sqrt{y_{\psi d}  \, y_{u b}}$  & $B_d \rightarrow \psiBbar \, n$ & 
 $3.6_{\pm 0.4} \! \cdot \! 10^{-5}$  \\
\, & &  $\,\,\,\, B_s \rightarrow \psiBbar \, \Lambda \,\,\,\,$ & n.a.  \\
\hline
 $\mathcal{O}_{s,ub}$  & $\sim 2.3 \sqrt{y_{\psi s}  \, y_{ub}}$ &  $\,\,B_d \rightarrow \psiBbar \, \Lambda \,\,$ & $1.3_{\pm 0.4} \! \cdot \! 10^{-4}$  \\  
\, & &  $B_s \rightarrow \psiBbar \,\Xi^0$ &  $2.0_{\pm 0.1} \!\cdot \! 10^{-5}$ \\
\hline
$\mathcal{O}_{d,cb}$ & $\sim 1.1 \sqrt{y_{\psi d}  \, y_{c b}}$  & $B_d \rightarrow \psiBbar \, \Sigma_c^0$ & $8.2_{\pm 0.4} \!\cdot \! 10^{-5}$ \\
\,& &  $B_s \rightarrow \psiBbar \, \Xi_c^0$& $7.0_{\pm 0.4} \!\cdot \! 10^{-5}$ \\
\hline
 $\mathcal{O}_{s,cb}$ & $\sim 1.1 \sqrt{y_{\psi s}  \, y_{c b}}$  & $B_d \rightarrow \psiBbar \, \Xi_c^0$ &$9.7_{\pm 5.0} \! \cdot \! 10^{-5}$  \\
\, & &  $B_s \rightarrow \psiBbar \, \Omega_c$ &  $1.3_{\pm 0.6} \!\cdot \! 10^{-4}$ \\ 
\hline

\end{tabular}
\caption{Analogous to Table \ref{tab:decays},  for operators with structure: $\mathcal{O}_{d_j, u_i b} \equiv i  \epsilon_{\alpha \beta \gamma} d_j^\alpha ( \bar{b}^{c \beta} u_i^\gamma).$
}
\label{tab:decays2}
\end{table}
\begin{figure*}[t!]
\centering
\setlength{\tabcolsep}{10pt}
\renewcommand{\arraystretch}{1}
\begin{tabular}{cc}
			\includegraphics[width=0.4\textwidth]{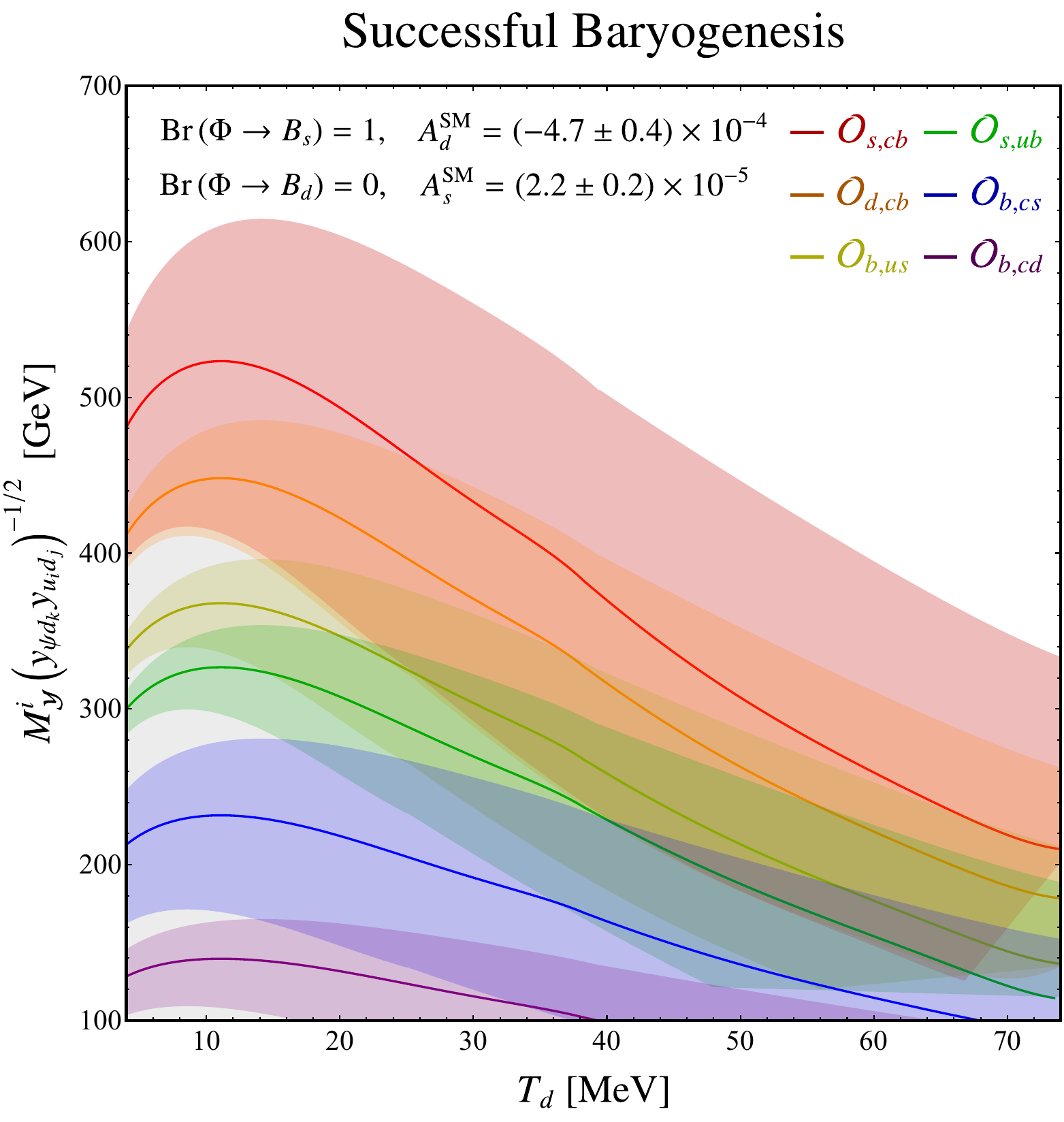} \,\,\,\,
&
		\,\,\,\, \includegraphics[width=0.4\textwidth]{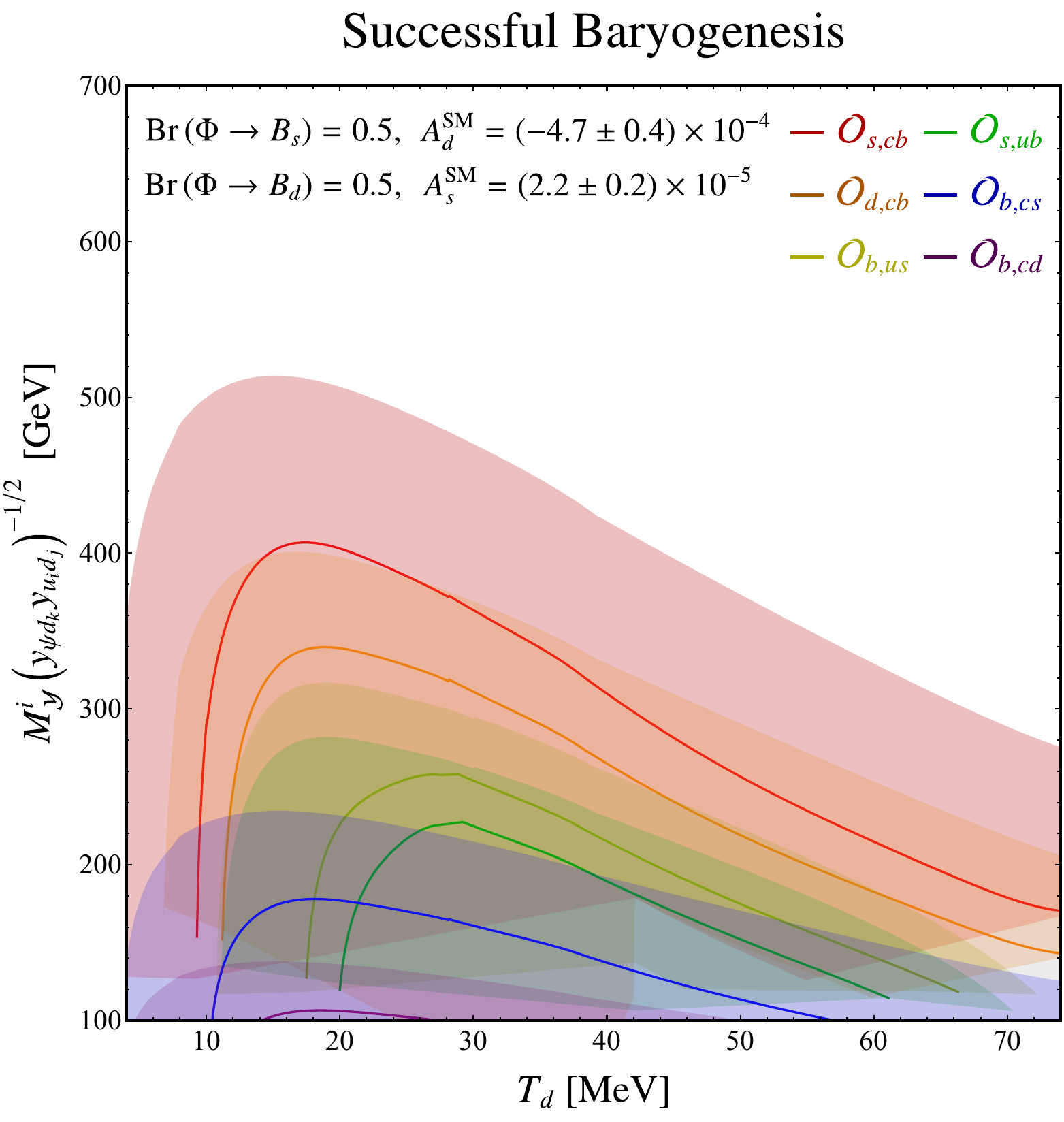}\\ \\ 
\end{tabular}
\vspace{-0.3cm}
\caption{Requisite initial $\Y$ mass and dark Yukawa couplings, such that the BAU is successfully generated through $\M$, with only the SM CPV in Eq.~\eqref{eq:SMCPV}. Solid lines correspond to the central value for each operator in Tables \ref{tab:decays}-\ref{tab:decays2}. Shaded bands account for a combination of uncertainties discussed in the text. Note that the entire parameter space below the upper line for each operator is allowed. In this sense, the upper curve corresponding to an upper bound on the maximal possible $M_\Y^i$. We require $M^i_\Y > 100 \, \text{GeV}$ to ensure that the operator Eq.\eqref{eq:op} is well defined.
\emph{Left:} The case where only $B_s$ mesons are produced in the early Universe, i.e. the branching fraction of $\Phi \rightarrow B_s$ is $100\%$. 
This case corresponds the maximal possible BAU generation (and smallest requisite $\Delta M_\Y$) with only the SM CPV. \emph{Right:} Scenario where $\text{Br}(\Phi \rightarrow B_s) = 0.5 = \text{Br} ( \Phi \rightarrow B_d)$. 
} 
\label{fig:MY}
\end{figure*}

From Eq.~\ref{eq:YBmeso}, it is clear that generating the observed BAU requires $\text{Br} \times A_{sl} \gtrsim 10^{-5}$. The SM CPV in $B^0$ oscillations have been calculated to be \cite{Lenz:2019lvd}:
\begin{subequations}\label{eq:SMCPV}
\begin{align}
&A_{sl}^d|_{\rm SM} = (-4.7 \pm 0.4 ) \times 10^{-4}  \,, \label{eq:aSMd}\\
& A_{sl}^s|_{\rm SM} = (2.1 \pm 0.2) \times 10^{-5}\,. \label{eq:aSMs}
\end{align}
\label{eq:SMCPV}
\end{subequations}
Meanwhile, recasting LEP constraints on $b$-quark decays constrains the Wilson coefficient $\mathcal{C}_{d_k, u_i d_j}$, and therefore the branching fraction $\Bri \lesssim 10^{-4}-10^{-3}$ \cite{Alonso-Alvarez:2021qfd}, as summarized in Tables \ref{tab:decays}-\ref{tab:decays2}.
Additionally, designated Mesogenesis searches by the Belle-II \cite{Belle:2021gmc} and BaBar \cite{BaBar:2023rer} collaborations have set limits on $B$-Mesogenesis through the $\psiB \, bus$ operator; $\text{Br}\left(B^0_d \rightarrow \Lambda + \text{MET} \right) \lesssim (2-4) \times 10^{-5}$. Clearly, the SM CPV, Eq.~\eqref{eq:SMCPV}, is not enough to generate the entire observed BAU.

We now make the observation that if the branching fraction $\Bri$ in the early Universe was enhanced compared to its present day value, the SM CPV could be enough to produce the observed BAU. This can be achieved if the mass of $\Y$ was temperature dependent, $M_\Y (T)$, such that $\Y$ was 
lighter during the epoch of Mesogenesis than today. In particular, we evoke the existence of some dark sector dynamics which \emph{morphs} the mediator mass  $M_\Y^i \rightarrow M_\Y^f > M_Y^i$ at a temperature $T_{\rm PT} < T_d \sim 5-80$ MeV. 
Thus present day collider constraints on $\Y$ are still maintained, but the branching fraction $\Br$ is enhanced in the early Universe relative to present day by a factor of $(M^f_\Y/M_\Y^i)^4$, lowering the requisite CPV to generate the BAU. Given the SM CPV in $B^0$-meson oscillations, it is clear from Eq.~\eqref{eq:YBmeso} that $\Br \gtrsim 3 \%$ for the SM CPV alone is sufficient to generate the entire BAU. Therefore, we expect that a mass shift of no more than $\Delta(M_\Y) \sim \mathcal{O}(\text{TeV})$ generates the BAU with only the SM CPV, whilst maintaining present day collider constraints. 

Detailed theoretical calculations of decay rates for $B_i \rightarrow \psiBbar \, \mathcal{B}_{\rm SM}$ for all possible SM final states were performed in \cite{Elor:2022jxy}. The decay rate goes as 
$\Gamma (B\rightarrow \psiBbar \, \mathcal{B}_{\rm SM}) = \left(y_{u_i d_j} y_{\psi d_k}/M_\Y\right)^4 \Gamma_0$, where  $\Gamma_0$ is a function of particle masses,
$\Delta m \equiv m_B - (m_{\psiB} + m_{\mathcal{B}_{\rm SM}}) \in [0,3 \, \text{GeV}]$. Maximal values of $\Gamma_0$, when $m_{\psiB} = 1$ GeV, are quoted in Tables \ref{tab:decays}-\ref{tab:decays2}. 
Fixing $A_{sl}^i$ to their SM values and expressing the branching fraction in terms of the $\Gamma_0$,   
\begin{align}
   \text{Br}_{B_i} = \frac{\sum_{\mathcal{B}_{\rm SM}} \mathcal{C}_i^2 \Gamma_0 (B_i \rightarrow \psiBbar \, \mathcal{B}_{\rm SM})}{(\tau_{B_{d,s}}^{\rm SM})^{-1} + \sum_{\mathcal{B}_{\rm SM}} \mathcal{C}_i^2 \Gamma_0 (B_i \rightarrow \psiBbar \, \mathcal{B}_{\rm SM})}~, \label{eq:BR}
\end{align} 
Eq.~\eqref{eq:YBmeso} is used to solve for $\mathcal{C}_{d_k,u_i d_j}$ such that the entire BAU is generated for a given operator in Tables \ref{tab:decays}-\ref{tab:decays2}. (The SM $B$ meson lifetime is $\tau_{B_{d,s}}^{\rm SM} \simeq 1.5\times 10^{-12}s$.) Results are shown in Fig.~\ref{fig:MY} where each colored band represents the needed $M_\Y^i$, with fixed Yukawas, such that the SM CPV alone generates the entire BAU. To obtain the requisite mass shift,  Fig.~\ref{fig:MY} can be compared to the bounds on the operator specific Wilson Coefficient in Tables \ref{tab:decays}-\ref{tab:decays2}. In much of the parameter space, a modest mass shift of 500 GeV$-$1 TeV can lead to successful baryogenesis through the \emph{3M} mechanism.

Given that the SM value  of $A_{sl}^d$ is negative,
$B_d^0$ decays will decrease the generated asymmetry, and as such \emph{3M} baryogenesis cannot proceed solely from $B_d^0$ mesons. The relative amounts of $B_s$ and $B_d$ produced in the early Universe depend on the model dependent fragmentation function of $\Phi$. To this end, we consider two cases in Fig.~\ref{fig:MY}: in the left panel we assume only $B_s$ mesons are produced, $\text{Br} (\Phi \rightarrow B_s) = 1$. Since $B_d$'s are not produced and do not deplete the asymmetry, this case represents the maximal possible enhancement to the early Universe branching fraction. In the right panel, we consider the case were $\Phi$ decays produce 50$\%$ $B_d$ and $B_s$ mesons \footnote{Note that  some fraction of $\Phi \rightarrow b\bar{b}$ could also fragment into $B^+$ as well. Results presented here can easily be extrapolated to account various scenarios.}; in this case the observed BAU can only be generated for higher values of $T_d$,  roughly above 15-20 MeV depending on the operator --- reflecting the effect of coherent $B_d$ oscillations (which peak at temperatures lower than for the $B_s$ system) depleting the asymmetry. In either case, for $T_d\gtrsim 70$ MeV, coherent oscillations are significantly suppressed due to electron scattering of the $B^0$ charge radius, and the baryon production ceases. 
Solid lines in Fig.~\ref{fig:MY} represent central values, while shaded regions corresponds to a combination of uncertainties from: the SM CPV values in Eq.~\eqref{eq:SMCPV}, the charge-radius of the neutral $B$ mesons (which translates into uncertainties in the decoherence functions $\alpha_{s,d}$ \cite{Alonso-Alvarez:2019fym, Alonso-Alvarez:2021qfd}), and  the QCD uncertainties in the branching fraction for each operator \cite{Elor:2022jxy}. Note that the bands are calculated using the maximal value of $\Gamma_0$, by fixing $m_{\psiB} = 1$ GeV. The entire area below each upper curve is allowed for larger values of $m_{\psiB}$.

\medskip
\mysec{Morphing with Dark Dynamics}
To generate the BAU with only the SM CPV in neutral $B_{s,d}$ meson oscillations, we must facilitate the \emph{morphing} of $\Y$'s mass such that it was light enough during the era of baryon production, but heavy enough (TeV scale) today to evade collider constraints.  As an example, consider a scenario in which $M_\Y$ depends on the vacuum expectation value (vev) of a scalar \emph{morphon} field $\phi$. The $\phi$ vev changes at a temperature $T_{\rm PT} \simeq 6-80\, \text{MeV}$, e.g. due to a delayed phase transition. For this scenario, we use the following morphon scalar potential: 
\bea
V_{\rm scalar} =  {m_\Y}_0^2 |\Y|^2 
    + y_{\phi \Y}  |\Y|^2 \phi 
    + \frac12\lambda_{\phi \Y} |\Y|^2 \phi^2\nonumber\\
    +  \frac14 \lambda ( \phi^2 - \phi_0^2 )^2 + \epsilon\, \phi_0 \,\phi^3\,.
\eea
The field-dependent mass of $\mathcal{Y}$ is;
\bea
M_{\Y}^2(\phi)
	&= m_{\Y_0}^2 + y_{\phi \mathcal{Y}} \phi + \frac12 \lambda_{\phi \mathcal{Y}} \phi^2 \,.
\eea
Note that $y_{\phi\Y}$ carries mass dimension one. Following the results displayed in Fig~\ref{fig:MY} and the constraints summarized in Tables~\ref{tab:decays}-\ref{tab:decays2}, we require that $M_\Y$ shifts between the false and true $\phi$-vacua such that:
\begin{subequations}
\begin{align}
M_\Y^i=M_\Y(v_{\rm false})
	&= \mathcal{O}(100 \text{ GeV})\,,
\\
M_\Y^f=M_\Y(v_{\rm true})
	&= \mathcal{O}(\text{TeV})\,.
\end{align}
\label{eq:MYshift}
\end{subequations}
\hspace{0.085in} For a successful $3M$ baryogenesis, the following general conditions are required of the morphon model: 
\emph{i. Nucleation}: The $\Y$ mass shift must occur after the BAU generation, i.e. $T_d > T_{\rm PT} > T_{\rm BBN} \sim 5 \, \text{MeV}$. Generically, in order to delay the vev change significantly below $T\sim \phi_0$, the effective potential for $\phi$ must have a high barrier between the minima and a small $\rho_{\rm vac}$, the energy difference between the true a false minimum. 
\emph{ii. Percolation}: The Universe must efficiently and completely transit from the false to the true morphon vacuum. 
\emph{iii. Avoiding Inflation}: 
The morphon, which remains trapped in a false vacuum until late times, must not dominate the energy density of the Universe triggering a period of inflation during or after the BAU is generated. 
Avoiding inflation selects morphon potentials with small $\rho_{\rm vac}$. 

In summary, $\M$ baryogenesis favors a fast percolating MeV scale phase transition, with a small $\rho_{\rm vac}$. While we leave open the possible models for generating such a potential, we present a specific example below.

\medskip

\begin{figure}[t!]
\centering
			\includegraphics[width=0.45\textwidth]{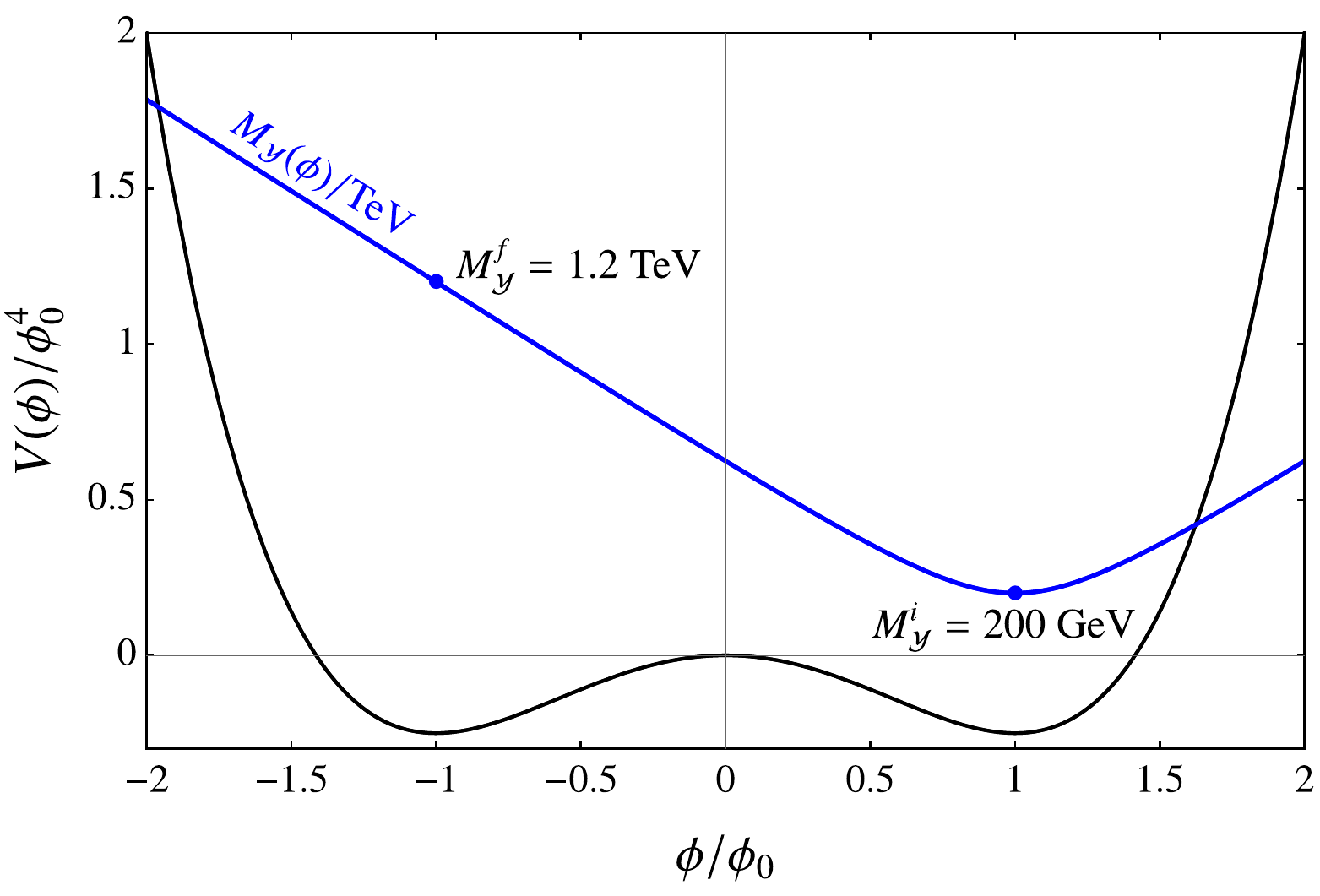}
\vspace{-.2cm}
\caption{The morphon potential $V(\phi)$ (black) that shifts the $\mathcal{Y}$ mass from $200\text{ GeV}$ in the false vacuum to $1.2\text{ GeV}$ in the true vacuum. In blue, we plot $M_\mathcal{Y}(\phi)$ to demonstrate the mass shift in $\mathcal{Y}$ from the false minimum $v_{\rm false} = \phi_0 + \mathcal{O(\epsilon)}$ to the (final) true minimum $v_{\rm true} = - \phi_0 + \mathcal{O(\epsilon)}$. In this plot, $\phi_0=10\text{ TeV},\ \lambda=1,$ and $\epsilon= -3\times 10^{-26}$. 
}
\label{fig:potential}
\end{figure}

\mysec{Example: Domain walls}
Consider the benchmark values:  $m_{\Y_0}=624.5\text{ GeV},\ y_{\phi \mathcal{Y}}=-70\text{ GeV},\ \lambda_{\phi \mathcal{Y}}= 0.007, \ \lambda=1, \ \epsilon=3\times 10^{-26}$ and $\phi_0=10\text{TeV}$ \footnote{It would be interesting to explore UV models in which $\epsilon$ can be made technically natural \cite{future}.}. The resulting potential and the $M_\Y$ shift is shown in Fig.~\ref{fig:potential}. This potential has nearly-degenerate minima at $T\simeq 20\text{ TeV}$. If the bias between the true and false vacuum is small enough, see Eq.~(\ref{eq:dw-perc}), the morphon field will fall into patches of true and false minima, forming a network of domain walls (DWs) that eventually annihilate~\cite{Gelmini:1988sf, Gelmini:2020bqg, Preskill:1991kd, Vilenkin:1981zs, Gleiser:1998na,Hiramatsu:2010yz, Kawasaki:2011vv, Hiramatsu:2013qaa}. 
DWs can grow to horizon size if Eq.~(\ref{eq:horizon}) is satisfied, separating large patches of the universe where $\phi$ has fallen into the false vacuum from large patches in the true vacuum $v_{\rm{false}/\rm{true}} = \pm \phi_0 + \mathcal{O(\epsilon)}$. 
The mass of $\mathcal{Y}$ is smaller in the false vacuum and larger in the true vacuum, see Fig.~\ref{fig:potential}. Since the BAU is mainly produced in patches where $\Y$ is light, i.e. the false vacuum, we need to overproduce the asymmetry initially. Therefore, for our example, we pick a false vacuum where $M_\Y \simeq 200$~GeV (well below the $M_\Y$ upper bounds shown in Fig.~\ref{fig:MY}).

Finally, we require that the DW network safely disappears. The DWs must annihilate by $T\sim 10 \text{ MeV}$, sufficiently before BBN, and the DWs must avoid dominating the energy density of the universe and triggering inflation. Relevant conditions on the evolution of the DW network are summarized in  the Supplementary Material.

\begin{figure}[t!]
\centering
\setlength{\tabcolsep}{10pt}
	\includegraphics[width=0.48\textwidth, trim=.1cm .2cm .1cm .0cm, clip=true]{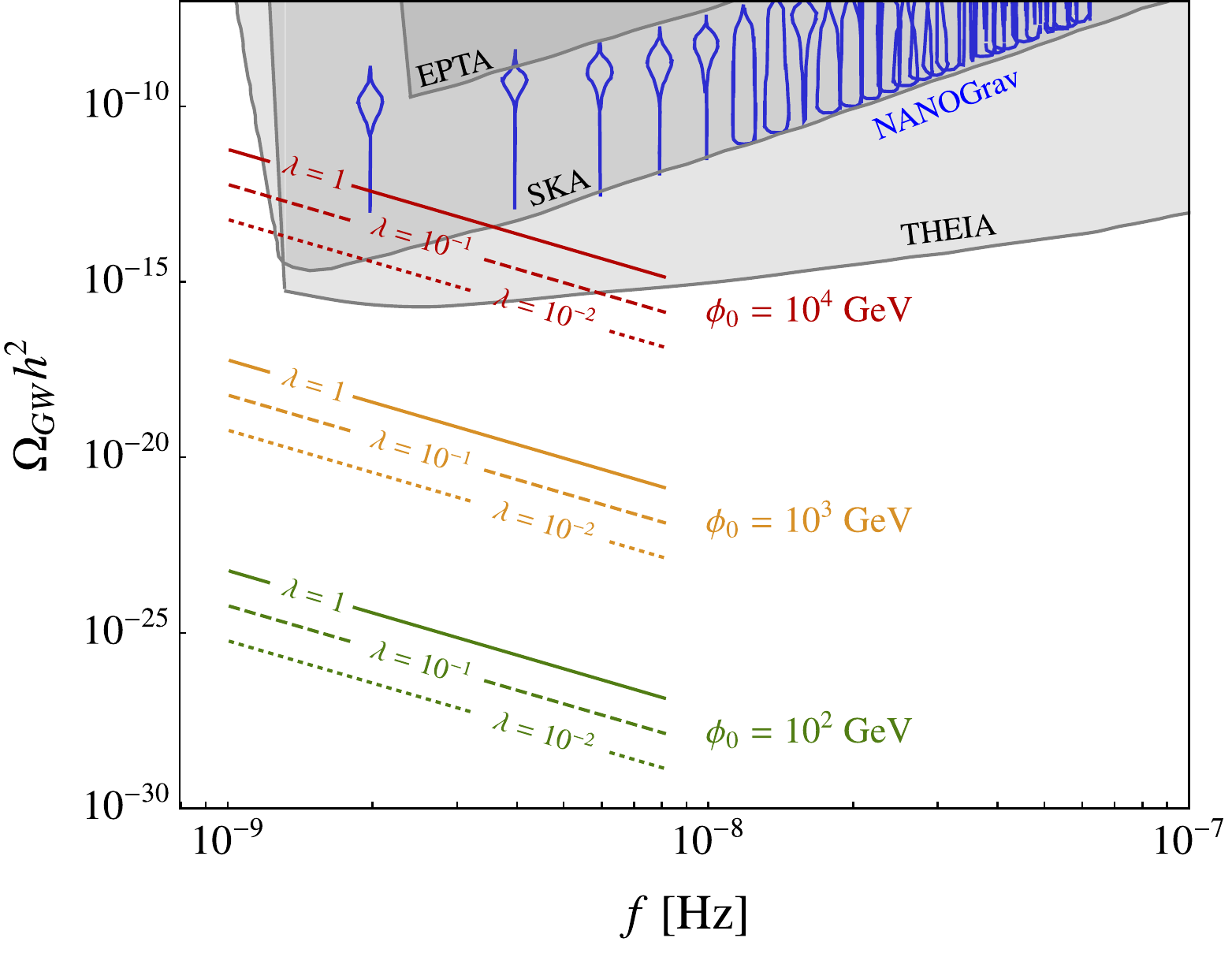}
\vspace{-0.4cm}
\caption{Each line corresponds to the peak of the gravitational wave power spectrum while scanning over the DW annihilation temperature, $T_{\rm ann}$, from $10$ MeV to $80$ MeV (corresponding to $f_{\rm peak}$ from $\sim1$ to $8$ nHz, respectively)  at fixed $\phi_0$ and $\lambda$. We also show current and future sensitivity curves for the European Pulsar Timing Array (EPTA) \cite{EPTA:2015qep}, the Square Kilometre Array (SKA) \cite{Janssen:2014dka}, GAIA/THEIA \cite{Garcia-Bellido:2021zgu} and NanoGrav \cite{NANOGrav:2023gor} \cite{NANOGrav:2023hvm}. 
} 
\label{fig:GW_signal}
\end{figure}

The annihilation of the DW network can leave behind a stochastic gravitational wave (GW) background~\cite{Riva:2010jm, Antusch:2013toa, Gelmini:2020bqg}. We used the expressions in Ref.~\cite{Gelmini:2020bqg} to calculate the GW peak frequency $f_{\rm peak}$ and the power spectrum $\Omega_{GW}$ evaluated at $f_{\rm peak}$. The temperature of the DW annihilation, $T_{\rm ann}$, sets the peak frequency of this GW signal. This mechanism requires the annihilation temperature for the DWs to be in the small window $10-80\text{ MeV}$, which translates to a prediction for the peak frequency of the GW signal power spectrum $f_{\rm peak}\in[1\text{-}8]\times10^{-9}\text{ Hz}$. Such frequencies are probed by Pulsar Timing Array (PTA) experiments~\cite{EPTA:2015qep, Janssen:2014dka,NANOGrav:2023gor, NANOGrav:2023hvm}.  We show the GW signal for a set of DW models that accomplish the desired mass shift of $\mathcal{Y}$ in Fig.~\ref{fig:GW_signal}.

\medskip
\mysec{Signals}
Violent processes in the early Universe, such as bubble collisions in a first-order phase transition,  
can source GWs. For the morphon scenario, the phase transition of interest will occur at late times, ($T_{\rm PT} \sim 10-50 \, \text{MeV})$, and be relativity fast corresponding to small values of the $\beta/H$ parameter. Such phase transitions are generically expected to yield low frequency GW signals that can be probed with current and future PTAs \cite{Schwaller:2015tja, EPTA:2015qep, Janssen:2014dka, Garcia-Bellido:2021zgu, NANOGrav:2023gor, NANOGrav:2023hvm}. 
This was demonstrated in the above DW example, but is expected to be a generic feature of dark dynamics that lead to the requisite morphing.
The details of the GW spectrum expected from $\M$ baryogenesis depend on the specific morphon model, and will be explored in upcoming work \cite{future}.

In the $\M$ baryogenesis, the branching fraction of the rare $B$ meson decays into SM baryons and missing energy measured at colliders today can be much smaller than the values predicted by $B$-Mesogenesis. Accordingly, a smaller branching fraction, $\text{Br} < 10^{-5}$, could be an indication that $\M$ with just the SM CPV is the mechanism responsible for generating the BAU and dark matter. This will be especially evident if PTAs see a GW peak consistent with morphing dark dynamics. The exact interplay between the GW signature and the predicted values of present day branching fractions will rely on the choice of the morphing model, which will be explored in detail in our companion paper \cite{future}.  Importantly, signals of the dark dynamics only complement the existing Mesogenesis search program. For instance, the dark dynamics do not alter the expected signal of induced proton decay in Mesogenesis \cite{Berger:2023ccd}.

\medskip
\mysec{Future Directions}
This letter introduced $\M$ baryogenesis which generates both the BAU and the dark matter abundance using SM CPV within $B$ meson systems.
We have therefore demonstrated that \emph{the Standard Model CP violation is, indeed, enough}. $\M$ can be tested through GW signals in conjunction with  collider searches already targeting Mesogenesis (but with better sensitivity), thereby paving the way to various experimental searches. Additionally, $\M$ opens up a variety of new theoretical directions.  While an explicit morphing model (in which DW annihilations lead to GW signals) was presented here, many possibilities, along with their signals, remain to be explored. For instance, morphing can be achieved through a first-order dark sector phase transition (delayed through supercooling or through the presence of a trigger field, e.g. \cite{Niedermann:2023ssr}). Such a model would be particularly interesting if the additional dark sector scalar fields could be realized in an inflationary context.  Another possibility is a  dark sector with multiple morphon fields, each obtaining a vev and contributing to $M_\Y$. Alternatively, multiple dark sector fields could lead to finite density effects in the dark sector which could morph the dark Yukawas instead of $M_\Y$ (the cumulative effect on the Wilson Coefficient is identical) in an analog to \cite{Fardon:2003eh}. Special care would need to be taken to ensure the number density of morphons does not trigger inflation. Yet another option would be to consider a dark confining phase transition with a large number of dark quark condensates contributing to $M_\Y$. The possible morphon models and their associated signals will be explored in our upcoming paper \cite{future}. Finally it would be interesting to construct models where the dark mediator $\Y$ is identified with a vector field.

\section*{Acknowledgments} 
\noindent 
The research of GE is supported by the National Science Foundation (NSF) Grant Number PHY-2210562, by a grant from University of Texas at Austin. SI is supported by the Natural Sciences and Engineering Research Council (NSERC) of Canada. RH was supported by the Institute for Fundamental Theory at
the University of Florida. MU is supported by a scholarship from Consejo Nacional de Ciencia Humanidades y Tecnologia (CONAHCYT) of Mexico. MU also acknowledges the CERN TH Department for hospitality while this research was being carried out.

\bibliography{Refs}

\clearpage

\onecolumngrid

\begin{center}
  \textbf{\large Supplemental Material for The Standard Model CP Violation is Enough}\\[.2cm]
  \vspace{0.05in}
  {Gilly Elor, \ Rachel Houtz, \ Seyda Ipek, \ and  Martha Ulloa Calzonzin}
\end{center}

\setcounter{equation}{0}
\setcounter{figure}{0}
\setcounter{page}{1}
\makeatletter
\renewcommand{\theequation}{S\arabic{equation}}

\section{Domain Wall Evolution and Constraints} 
\label{sec:DW-evolution}

To accomplish baryogenesis, we require that the four-fermion operator responsible for translating the CP violation in $B$-meson oscillations to a baryon asymmetry between the visible and dark sectors must be enhanced. This is accomplished via a phase transition that morphs the mediator $\mathcal{Y}$ from a lighter mass in the false vacuum to a heavier mass in the true vacuum. One example presented here is a phase transition that proceeds via rolling from a symmetry-preserving minimum into two sets of nearly degenerate minima. One minimum is the false vacuum, with a small $\mathcal{Y}$ mass, and the other is a true vacuum with a large $\mathcal{Y}$ mass. In our example, a domain wall network forms between patches of the universe with different $\phi$ vevs. 

To ensure our DW-forming potential accomplishes baryogenesis and evades current constraints, we put a few requirements on the potential parameters, following the analysis of~\cite{Gelmini:2020bqg}. First, the two minima must be degenerate enough to reach a percolation threshold and form a DW network initially. Second, our mechanism requires large regions of the universe to be in the false vacuum during baryogenesis, meaning the walls must reach horizon size. Third, the DW network must disappear at $T\geq 10\text{ MeV}$ to safely avoid spoiling BBN observations. Finally, the energy stored in the domain walls must not dominate the energy density of the universe and trigger inflation. 

There are two competing parameters that determine the evolution of the DW network: the vacuum pressure and the surface pressure of the walls. The vacuum pressure $p_{\rm vac} = \epsilon \phi_0^4$ encodes the preference for the true vacuum over the false vacuum, pushing regions of true vacuum to expand and accelerating the walls until they annihilate. The surface pressure is $p_T = \sigma / R$, where $\sigma = (2\sqrt{2}/3)\sqrt\lambda \phi_0^3 $ is the surface tension and $R$ is the average radius of the curvature of the wall. Roughly, the surface pressure decreases when the walls grow and flatten out. The walls annihilate when $p_T>p_{\rm vac}$. This means our first two conditions, which require the percolation and growth of DWs, will put an upper bound on the vacuum degeneracy encoded in $\epsilon$. Our last two constraints, which require a disappearance of DWs, will put a lower bound. 

The constraints we put on our potential are the following:
	\begin{align}
	\epsilon 
		&\lesssim 0.2 \lambda
			&& \text{ DWs percolate~\cite{Gelmini:1988sf, Stauffer:1978kr} }
        \label{eq:dw-perc}
	\\
    	\epsilon	
	    	&< \frac{2 \sqrt2 }3
   	 		\sqrt{\frac{ 8\pi^3 g_{\rm eff}}{90}} \frac{T^2}{ M_{Pl}}
     			\frac{ \sqrt\lambda }{\phi_0}  \,\bigg|_{T = T_c = 2 \phi_0 }
			&& \text{ DWs grow to horizon size, $p_{\rm vac} < p_T$, when $R(T) \sim 1/2H(T)$}
	\label{eq:horizon}
	\\
	\epsilon
		&> \frac{2 \sqrt2 }3
   	 		\sqrt{\frac{ 8\pi^3 g_{\rm eff}}{90}} \frac{T^2}{ M_{Pl}}
     			\frac{ \sqrt\lambda }{\phi_0}  \,\bigg|_{T = 10\text{ MeV } }
			&& \text{ DWs annihilate, $p_{\rm vac} > p_T$,   at $10\text{ MeV}$}
	\label{eq:bbn}
	\\
	\epsilon
		&> \left( \frac43 \right)^3 \frac{ 4\pi \lambda \phi_0^2 }{M_{Pl}^2 }
			&& \text{ DWs annihilate before they trigger inflation}\,,
	\label{eq:infl}
	\end{align}
where $g_{\rm eff}$ is the effective active number of degrees of freedom, $M_{Pl}$ is the Planck mass, and $T_c$ is the critical temperature near which the domain walls form, estimated to be $T_c\sim2\phi_0$. The constraints in Eq.~(\ref{eq:horizon}) and~(\ref{eq:bbn}) have the same form because once the walls reach horizon size, they enter a scaling regime and continue to follow $R \sim 1/2H$. To arrive at Eq.~(\ref{eq:infl}), we require that the walls annihilate before the walls trigger inflation, again closely following the analysis of Ref.~\cite{Gelmini:2020bqg}. The walls trigger inflation when the energy density stored in the walls, $\rho_w \sim \sigma/R$, exceeds the critical density $\rho_c = 3 M_{Pl}^2/32 \pi t^2$. The time when inflation would begin must be larger than the time when the walls annihilate, $t_{\rm ann} \sim \sigma/\epsilon \phi_0^4$, yielding Eq.~(\ref{eq:infl}).

\label{app:}

\twocolumngrid

\end{document}